\documentclass{article}
\usepackage[brazil]{babel}
\usepackage[latin1]{inputenc}
\newcommand{\p}{\partial}
\newcommand{\al}{\alpha}
\newcommand{\be}{\beta}
\newcommand{\ep}{\epsilon}
\newcommand{\de}{\delta}
\newcommand{\ti}{\tilde}

\usepackage{indentfirst}
\title{Generalization of the Extended Lagrangian Formalism on a Field
Theory and Applications}
\author{A. A. Deriglazov\footnote{alexei.deriglazov@ufjf.edu.br} \and B. F.
Rizzuti\footnote{brunorizzuti@ufam.edu.br ~ On leave of absence
from Instituto de Sa\'ude e Biotecnologia, ISB, Universidade
Federal do Amazonas, Coari-AM, Brazil}}

\date{Dept. de Matem\'atica, ICE, \\
Universidade Federal de Juiz de Fora, MG, \\
and \\
Dept. de F\'isica, ICE, \\
Universidade Federal de Juiz de Fora, MG}

\begin{document}
\maketitle \large

\begin{abstract}
Formalism of extended Lagrangian represent a systematic procedure
to look for the local symmetries of a given Lagrangian action. In
this work, the formalism is discussed and applied to a field
theory. We describe it in detail for a field theory with
first-class constraints present in the Hamiltonian formulation.
The method is illustrated on examples of electrodynamics,
Yang-Mills field and non-linear sigma model.
\end{abstract}

\noindent
{\bf PAC codes:} 11.30.-j, 11.10.Ef 

\section{Introduction}
In the field theory with local symmetries, the number of variables
used in the description is greater than the number of degrees of
freedom. It is important to keep all the variables used to
guarantee, for instance, the manifest Lorentz covariance. On the
other hand, one needs to characterize, in one way or another, the
physical sector of a given theory. This can be achieved using the
manifest form of the local symmetry: among all variables, the
physical ones turn out to be invariant under the action of local
symmetries. So, knowledge the local symmetries in many cases is
crucial in analysis of physical content of a theory.

The locally invariant theories are described by singular
Lagrangians, so their analysis is carried out in accordance with
the Dirac method for constrained systems [1]. The presence of
constraints in the Hamiltonian formulation reflects the fact that
the dynamics of part of the variables is dependent on the
remaining ones. The constraints are divided into two groups: the
first class and the second class. It is well known that the
first-class constraints are closely related to local symmetries
[2, 3, 4]. So, an interesting problem under investigation by
various groups [5-22] is whence there is a relatively simple and
practical procedure for restoration the symmetries from the known
constraints. In the Hamiltonian formalism, the problem has been
solved for the case of mechanical system with first-class
constraints along the following line [3]. The initial Hamiltonian
action (which by construction contains the primary constraints
only) can be replaced on the extended Hamiltonian action, with all
the higher-stage constraints with their own Lagrangian multipliers
added to the action. It leads to the equivalent formulation [2].
Local symmetries of the extended Hamiltonian action have been
found in the closed form [3]. Moreover, in absence of second-class
constraints, local symmetries of the initial Hamiltonian action
can be restored in the algebraic way [3].

Search for the local symmetries of the initial Lagrangian action
represents a separate issue. for the mechanical constrained system
with first and second class constraints, one possible way to solve
the problem has been developed in the works [5, 6]. Given a
singular Lagrangian $L$, the theory can be reformulated in terms
of an extended one, $\tilde L$, equivalent to $L$. Due to special
structure of $\tilde L$, its gauge symmetries can be found in a
closed form. All the first class constraints of $L$ turn out to be
the gauge generators of the symmetries of $\tilde L$. The extended
Hamiltonian of initial theory turns out to be the Hamiltonian for
the extended Lagrangian [6]. For a theory with first-class
constraints, it is also possible to find the symmetries of the
initial Lagrangian $L$ [3, 5, 6]. The aim of this work is to
discuss the method described above to the case of a field theory,
showing explicitly the differences that arise when we move on from
mechanical to a field theory, and apply it to particular models. 

In [5, 6] we consider an action invariant modulo the total
derivative term. It should be mentioned that by appropriately
extending an action, one can made it exactly invariant [7-11]. The
modified action contains a surface term which is, in general,
different from zero [7, 8] (the generalizations for the field
theory and to arbitrary or noncanonical symplectic structures may
be found in [9, 10]). In this case analysis of the Hamiltonian
action shows that the Hamiltonian generators acquire the surface
term [11]. The method turns out to be useful in the path-integral
quantization framework of a generally covariant theories in
time-independent gauges [11].

This paper is divided as follows. Section 2 is devoted to discuss
the method of finding local symmetries for singular mechanical
models. In Section 3 we generalize this method for constrained
field models. The method is illustrated on the examples of
electrodynamics, Yang-Mills field and non-linear sigma model on
Section 4. Section 5 is left for conclusions.

\section{Search for symmetries - Extended Lagrangian approach}

This section is devoted to review the method of finding local
symmetries of a singular mechanical system [4, 5, 6]. It is done
by deforming the initial Lagrangian in such a way that all its
symmetries can easily be found in closed form. As it will be
shown, all the first class constraints of the initial Lagrangian
turn out to be the gauge generators of local symmetries of the
deformed Lagrangian. The symmetries of the initial Lagrangian are
also found.

\subsection{Construction of extended Lagrangian and Hamiltonian}

Starting from a singular Lagrangian $L(q^A, \dot q^A)$, one
applies the Dirac procedure, obtaining Hamiltonian and complete
Hamiltonian given by $H_0$ and $H$. The system of constraints is
given by $\{G_I \}=\{\phi^{\al}, T_a\}$, where $\phi^{\al}$ are
primary constraints and we denote $T_a$ all the further stage
constraints. We suppose that all of them are first class (they
obey the algebra $\{G_I,G_J\}=c_{IJ}{}^K G_K$, $\{G_I,
H\}=b_I{}^JG_J$) and that the procedure stops at $N$-th stage. It
is equivalent to the existence of local symmetries for $L$ of the
type [2, 3, 12],
\begin{equation}\label{11.11}
\delta q^A= \varepsilon R^A_0 + \dot \varepsilon R^A_1 + ...+
\frac{d^{N-1}\varepsilon}{d \tau^{N-1}} R^A_N.
\end{equation}

We construct the following function, defined on phase space
parameterized by $q^A, \tilde p_A, s^a, \pi_a, v^{\al}, v^a$,
\begin{eqnarray}\label{13}
\tilde H(q^A, \tilde p_A, s^a, \pi_a, v^\alpha, v^a)=\tilde
H_0(q^A, \tilde p_j, s^a)+v^\alpha\phi_\alpha(q^A, \tilde
p_B)+v^a\pi_a,
\end{eqnarray}
where,
\begin{eqnarray}\label{14}
\tilde H_0=H_0(q^A, \tilde p_j)+s^aT_a(q^A, \tilde p_j).
\end{eqnarray}
The functions $\phi_{\al}$, $H_0$ and $T_a$ were taken from the
initial formulation.

We affirm that $\ti H$ is the complete Hamiltonian for a
Lagrangian $\ti L(q^A, \dot q^A, s^a)$ (to be determined), $\ti
H_0$ is the Hamiltonian for $\ti L$ as well as $\phi_{\al}=0$ and
$\pi_{a}=0$ are primary constraints ($\pi_a$ are conjugate momenta
for $s^a$ variables). Furthermore, $L$ and $\ti L$ are equivalent.
To show all these facts, first we write the following equation of
motion,
\begin{eqnarray}\label{15}
\dot q^i=\frac{\partial\tilde H}{\partial\tilde p_i}=
\frac{\partial H_0}{\partial\tilde p_i}- v^\alpha\frac{\partial
f_\alpha}{\partial\tilde p_i}+s^a\frac{\partial
T_a}{\partial\tilde p_i}.
\end{eqnarray}
This equation can be inverted with respect to $\ti p_i$ in a
neighborhood of the point $s^a=0$ (for details, see [5]). Let us
denote the solution as,
\begin{eqnarray}\label{16}
\tilde p_i=\omega_i(q^A, \dot q^i, v^\alpha, s^a).
\end{eqnarray}
Now, on space $q^A$,$s^a$ we define,
\begin{eqnarray}\label{17}
\tilde L(q^A, \dot q^A, s^a)= \qquad \qquad \qquad \qquad \qquad
\cr \left.\left(\omega_i\dot q^i+f_\alpha(q^A, \omega_j)\dot
q^\alpha- H_0(q^A, \omega_j)-s^aT_a(q^A,
\omega_j)\right)\right|_{\omega_i(q, \dot q, s)}.
\end{eqnarray}
In the definition above we have used the notation,
\begin{eqnarray}\label{18}
\left.\omega_i(q^A, \dot q^i, v^\alpha,
s^a)\right|_{v^\alpha\rightarrow\dot q^\alpha}\equiv\omega_i(q,
\dot q, s).
\end{eqnarray}

If we now suppose that $\ti L$ is some singular Lagrangian, then a
direct calculation shows that $\ti H_0$ and $\ti H$ are its
corresponding Hamiltonian and complete Hamiltonian, respectively.
The Dirac method applied to $\ti H$ shows that all the higher
stage constraints of the initial theory are now, at most,
secondary ones. It implies, in particular, that the local symmetry
of $\ti L$ is of $\dot \epsilon$-type, and hence has simple
structure as compared to ${\stackrel{(N-1)}{\epsilon}}$-type
symmetry of initial formulation (see Eq. (\ref{11.11})). If now
one fixes the gauge $s^a=0$ for the constraints $\pi_a=0$, the
sector $(s^a,\pi_a)$ disappears of the extended formulation. Then
one is faced again with the initial formulation. Since $L$ is one
of the gauges of $\ti L$, the equivalence between the two
formulations is proved. Hence, it is only matter of convenience to
analyze the extended or the initial Lagrangian.

\subsection{Restoration of local symmetries}

Before we obtain the local symmetries of extended and initial
forumlation, it is important to note two points, already cited in
Introduction. The first one is that a gauge symmetry, in
Lagrangian or Hamiltonian actions, is defined modulo a total
derivative. Moreover, we want to find local symmetries of the
initial Lagrangian action. These two topics make our analysis
different from the one considered in the papers [7, 8], where the
main idea is to reformulate only the Hamiltonian action, adding
boundary terms, to make it fully gauge-invariant. We start with
the extended Hamiltonian formulation, passing through extended
Lagrangian action and finally we arrive at the initial
formulation.

We will begin with the Hamiltonian action,
\begin{eqnarray}\label{r.1}
S_{\tilde H \tilde L}=\int d\tau(\tilde p_A\dot q^A+\pi_a\dot
s^a-\tilde H).
\end{eqnarray}
According to Dirac conjecture [3], the first class constraints are
believed to generate gauge transformations. So, one considers the
transformations $\delta_I q^A=\epsilon^I\{q^A, G_I\}$, ~
$\delta_I\tilde p_A=\epsilon^I\{\tilde p_A, G_I\}$, where
$\epsilon^I=\epsilon^I(\tau)$ are arbitrary functions, that is not
necessarily zero at the endpoints and $I$ may assume any fixed
value $\alpha$ or $a$. Omitting total derivative terms, it is
possible to show that these transformations implies that $\de
S_{\tilde H \tilde L}$ is proportional to $\phi_\alpha, T_a$.
Then, it is possible to find appropriate transformations for
$v^\alpha, s^a$, that leaves $S_{\tilde H \tilde L}$ invariant. In
fact, direct calculations shows that the transformations below,
\begin{eqnarray}\label{r.2}
\delta_I q^A=\epsilon^I\{q^A, G_I\}, \qquad \delta_I\tilde
p_A=\epsilon^I\{\tilde p_A, G_I\}, \qquad \qquad \quad ~ ~ \cr
\delta_I s^a=\dot\epsilon^a\delta_{aI}+\epsilon^I
b_I{}^a-s^b\epsilon^I c_{b I}{}^a- v^\beta\epsilon^I c_{\beta
I}{}^a, \qquad \delta_I\pi_a=0, \cr \delta_I
v^\alpha=\dot\epsilon^\alpha\delta_{\alpha I}, \qquad \delta_I
v^a=(\delta_I s^a)^{.} \qquad \qquad \qquad \qquad \qquad ~ ~
\end{eqnarray}
keep the Hamiltonian action invariant (modulo a surface term) [5].
It prompts us to find the symmetries of the extended Lagrangian
action,
\begin{equation}\label{19}
S_{\ti L}=\int d \tau \ti L,
\end{equation}
in closed form. Namely the following variations
\begin{eqnarray}\label{20}
\delta_I q^A=\epsilon^I\left.\{q^A,
G_I\}\right|_{p\rightarrow\omega(q, \dot q, s)}, ~\Leftrightarrow
~ \left\{
\begin{array}{ccc}\label{20.1}
\delta_I q^\alpha & = & \epsilon^\alpha\delta_{\alpha I},\\
\delta_I q^i & = & \epsilon^I\left.\frac{\partial
G_I}{\partial\tilde p_i}
\right|_{p\rightarrow\omega(q, \dot q, s)};\\
\end{array}
\right. \cr \delta_I
s^a=\left.\left(\dot\epsilon^a\delta_{aI}+\epsilon^I
b_I{}^a-s^b\epsilon^I c_{b I}{}^a- \dot q^\beta\epsilon^I c_{\beta
I}{}^a\right)\right|_{p\rightarrow\omega(q, \dot q, s)}. \qquad
\qquad  ~ ~
\end{eqnarray}
represent the local symmetries of the action. This demonstration
may be found in [5].

Let us obtain the symmetries of the initial action. To do this, we
must eliminate the sector $s^a$ of the extended formulation in an
appropriate way. So, consider the combination of symmetries of
$\ti L$,
\begin{equation}\label{21}
\delta \equiv \sum_I \delta_I,
\end{equation}
which obeys $\delta s^a=0$ for all $s^a$. If one uses the property
$\ti L(q^A, \dot q^A, s^a=0)=L(q^A, \dot q^A)$, then $L$ is
invariant under any transformation,
\begin{equation}\label{22}
\delta q^A=\sum_I \delta_I q^A \Big|_{s^a=0},
\end{equation}
which obeys $\delta s^a=0 \Big|_{s^a=0}$, that is,
\begin{equation}\label{23}
\dot \epsilon^a+\epsilon^I b_I{}^a-\dot q^{\be}c_{\be I}{}^a=0.
\end{equation}
We have $[a]$ equations for $[\al]+ [a]$ variables $\epsilon^I$.
When there are only first class constraints, this system can be
solved iteratively [3], leading to $[\al]$ local symmetries of
$L$. This cumbersome calculation is given in [5]. We observe that
we are not discarding surface terms. They are absorbed in the
definition of gauge transformation in both cases: Hamiltonian and
Lagrangian actions.

In the presence of second class constraints, local symmetries of
$L$ can not be generally restored according to the procedure
discussed above. The reason is that a number of equations of the
system (\ref{23}) can be equal or more than the number of
parameters $\epsilon^a$, see an example of this kind in the work
[6].

\section{Gauge symmetries for constrained field models}

Let us discuss the method of finding local symmetries for
constrained field models. It will be carried out in the same way
as described in the previous section. However we will point out
some special novelties which are present when the method is
applied for a singular field model.

Let we have a singular Lagrangian $L=\int d^3 x \mathcal
L(\varphi^A, \partial_{\mu}\varphi^{A})$. The indices $A$ may
correspond to various types of fields. The conjugate momenta are
defined by
\begin{equation}\label{g.1}
p_A=\frac{\de L}{\de \dot \varphi^A}=\frac{\p \mathcal L}{\p \dot
\varphi^A}.
\end{equation}
Suppose that we have carried out the corresponding
hamiltonization. The notation follows directly from the previous
Section. Since $L$ may depend on spatial derivative of the fields,
we observe that further stage constraints may depend on spatial
derivative of the momenta. It gives rise to the first novelty when
we begin the procedure of finding local symmetries. We write the
equation of motion,
\begin{equation}\label{23.1}
\dot \varphi^i =\frac{\p \mathcal H_0}{\p \ti p_i}-v^{\al}\frac{\p
f_{\al}}{\p \ti p_i}+ s^a\frac{\de T_ a}{\de \ti p_i}.
\end{equation}
This equation should be inverted in terms of $\ti p_i$ to
construct the extended Lagrangian. Nevertheless, in general case
one is faced with a partial derivative of $\ti p_i$. To avoid this
problem, let us suppose that the constraints are, at most, linear
in spatial derivative of the momenta. In this case, Eq.
(\ref{23.1}) can be inverted. We point out that constraints with
polynomial form in fields and corresponding momenta do not
represent any restriction to inversion of (\ref{23.1}), see [5].
Although restrictive, to our acknowledge, all important physical
models that possess local invariance bear this particular
structure in hamiltonian formulation, \emph{i.e.} with linear
constraints in spatial derivative of the momenta. Indeed,
Electrodynamics, Yang-Mills field, Standard Model, string and
membrane theories are of this type. There is another novelty that
must be taken into account: the coefficients of the gauge algebra
may be not functions but operators, \emph{e.g.}, $\{G_I,G_J\} \sim
\p_j\p^j G_K$. Finally, the gauge generators are,
\begin{equation}\label{23.2}
G=\int d^3x \epsilon^I(x)G_I(x).
\end{equation}
Integration is taken over all the space. The method of finding
symmetries is now carried out analogously.

At this point, it may be interesting to discuss certain special
subtleties present in singular field models that do not result
directly from the generalization of point mechanics to the
continuous case. In classical systems, physical degrees of freedom
are understood to be the minimum number of variables necessary to
fully describe the model. In a field theory, physical degrees of
freedom can be understood to be the minimum number of fields in
each point of underlying space where the fields are defined, which
completely describe the model. For instance, we say that
Electrodynamics has 2 degrees of freedom, since it is possible to
eliminate 2 of the 4 components of the vector
$A_{\mu}=A_{\mu}(x)$, for each point in space-time parameterized
by $x^{\mu}$. In fact, there are 8 field components in phase space
($A_{\mu}$ and its corresponding momenta) together with 2 first
class constraints. After the gauge is fixed, we are left with 4
second class constraints. Hence there are only 8-4=4 independent
field components. Consequently, only 2 components of $A_{\mu}$
remain independent on configuration space of fields. We must also
be careful with the meaning of constraint in a field theory. In
classical systems, each constraint (algebraic equation involving
coordinates and momenta) allows us to eliminate one differential
equation from all the equations of motion describing the model.
This means that not all variables have independent dynamics. For a
field theory, a constraint can also be a differential equation.
Hence elimination of non-physical degrees of freedom does not
follow directly. This point may also be exemplified using
Electrodynamics: $p_i$ are the conjugate momenta for $A_i$ and
$p_0 \approx 0$ is the primary constraint. The evolution of $p_0$
leads to the secondary constraint $\p_i p_i \approx 0$. The
elimination of any degree of freedom using the secondary
constraint is not as obvious as it is for the primary one. (For a
discussion of the above points, see [13, 14, 23]).

\section{Applications}

We will consider some specific examples of constrained field
models for applying the method presented, including
Electrodynamics, Yang-Mills and nonlinear sigma model.

\subsection{Local symmetry of electrodynamics}

Let us consider the Lagrangian of electromagnetic field,
\begin{equation}\label{26}
L=-\frac{1}{4} \int d^3 x F_{\mu \nu} F^{\mu \nu}=\int d^3 x
[\frac{1}{2}(\dot A_i-\p_iA_0)^2-\frac{1}{4}F_{ij}F^{ij}],
\end{equation}
where $F_{\mu \nu}=\p_{\mu}A_{\nu}-\p_{\nu}A_{\mu}$. The primary
constraint and conjugate momenta are given by,
\begin{equation}\label{27}
\frac{\p \mathcal L}{\p \dot A_0}=p_0=0 \Rightarrow \phi_1 \equiv
p_0 =0,
\end{equation}
\begin{equation}\label{28}
\frac{\p \mathcal L}{\p \dot A_i}=p_i=\dot A_i-\p_iA_0 \Rightarrow
\dot A_i=p_i+\p_iA_0.
\end{equation}
The Hamiltonian $H_0$ and complete Hamiltonian $H$ are,
\begin{equation}\label{29}
H_0= \int d^3 x [\frac{1}{2}p_i^2+p_i\p_iA_0+\frac{1}{4}F_{ij}^2],
\end{equation}
\begin{equation}\label{30}
H=H_0+\int d^3 x v^0p_0,
\end{equation}
where $v^0$ is the corresponding Lagrange multiplier. The
secondary constraint follows from the consistency condition
$0=\{p_0(x_1),H\}$. It leads to $T_2 \equiv \p_i p_i=0$. There are
no further constraints.

The gauge algebra is,
\begin{eqnarray}\label{32}
\{p_0, \p_i p_i\}=0, \, \{p_0,H_0\}=\p_i p_i, \, \{\p_i
p_i,H_0\}=0.
\end{eqnarray}
The extended Hamiltonian takes the form,
\begin{equation}\label{35}
\ti H=\int d^3 x\Big[ \frac{1}{2}\ti p_i^2 + \ti p_i \p_i A_0 +
\frac{1}{4}F_{ij}^2 + s^2\p_i \ti p_i+ v^2 \pi_2 + v^0 \ti p_0
\Big].
\end{equation}
Starting from,
\begin{equation}\label{36}
\dot A_i=\{A_i,\ti H\}=\ti p_i +\p_i A_0-\p_is^2 \Rightarrow \ti
p_i=\dot A_i -\p_i A_0 +\p_is^2,
\end{equation}
we find $\ti L$,
\begin{equation}\label{37}
\ti L=\int d^3 x\Big[ \frac{1}{2}(\dot A_i
-\p_iA_0+\p_is^2)^2-\frac{1}{4} F_{ij}^2 \Big].
\end{equation}
The symmetries of $\ti L$ are given by,
\begin{equation}\label{38}
\de_1: \,\, \de_1A_i=0; \, \de_1A_0=\de_1s^2=\ep^1,
\end{equation}
\begin{eqnarray}\label{39}
\de_2: \,\, \de_2 A_i=\int d^3x \ep^2
\{A_i,\p_l p_l\}=-\p_i \ep^2, \\
\de_2 A_0 = 0; \, \de_2 s^2= \dot \ep^2.
\end{eqnarray}
The symmetries of $L$ are directed restored, see (\ref{22}) and
(\ref{23}),
\begin{eqnarray}\label{40}
\de_1A_i+\de_2A_i=-\p_i \ep^2, \\
\de_1A_0+ \de_2A_0 = \ep^1,
\end{eqnarray}
where the $\ep$'s obey the equation,
\begin{equation}\label{41}
\dot \ep^2 + \ep^1=0 \Rightarrow \ep^1=-\dot \ep^2.
\end{equation}
Defining $\ep^2 \equiv - \al$, we obtain the well known gauge
symmetry of electrodynamics,
\begin{equation}\label{42}
A_{\mu}(x^{\nu}) \rightarrow A'_{\mu}(x^{\nu})= A_{\mu}(x^{\nu})+
\p_{\mu}\al(x^{\nu}),
\end{equation}
where $\al=\al(x^{\nu})$ is an arbitrary space-time scalar
function.

\subsection{Local symmetry of Yang-Mills field}

In the pioneer work [24], Yang and Mills (YM) have considered the
idea of interact a original set of fields, invariant under a group
with constant parameters, with a new field (gauge field). It was
accomplished by postulating the invariance of the system under the
original group but having now arbitrary functions as parameters.
We will discuss this field model via Dirac procedure and we shall
find its local symmetries. Let us consider the YM Lagrangian,
\begin{equation}\label{43}
L=\int d^3x \mathcal L=-\frac{1}{4} \int d^3xF^{a}_{\mu \nu}F^{a
\mu \nu},
\end{equation}
where $F^a _{\mu \nu}= \p_{\mu}
A^{a}_{\nu}-\p_{\nu}A^{a}_{\mu}+igf^{abc}A^{b}_{\mu}A^{c}_{\nu}$.
$L$ has global $SU(N)$ symmetry, the field $A_{\mu}$ assumes
values on the corresponding Lie algebra with generators $T^a$,
\begin{equation}\label{44}
A_{\mu}=A^{a}_{\mu}T^a,
\end{equation}
and $f^{abc}$ are the structure constants,
\begin{equation}\label{45}
[T^a,T^b]=if^{abc}T^c.
\end{equation}
The primary constraints and conjugate momenta are,
\begin{eqnarray}\label{46}
\frac{\p \mathcal L}{\p \dot A^{a}_0}=p^a_0=0 \Rightarrow
\phi^a_1=p^a_0=0, \\
\frac{\p \mathcal L}{\p \dot A^a_i}=p^a_i=\dot A^a_i-\p_iA^a_0+ig
f^{abc}A^b_0 A^c_i \Rightarrow \cr \dot A^a_i=p^a_i + \p_iA^a_0-ig
f^{abc}A^b_0 A^c_i.
\end{eqnarray}
The Hamiltonian $H_0$ and complete Hamiltonian $H$ are given by,
\begin{eqnarray}\label{47}
H_0=\int d^3x \Big[\frac{1}{2} (p^a _i)^2 + p^a _i \p_iA^a_0 -
igf^{abc}A^b_0 A^c_i p^a_i+ \frac{1}{4}(F^a_{ij})^2 \Big] \\
H=H_0+ \int d^3x \lambda^a p^a _0,
\end{eqnarray}
where $\lambda^a$ are the corresponding Lagrange multipliers. The
secondary constraint follows from the consistency condition
$0=\{p^a_0(x_1),H\}$. One finds, $T^a_2=\p_i p^a_i-igf^{abc}p^b_i
A^c_i=0$. There are no further constraints. The gauge algebra is,
\begin{eqnarray}\label{48}
\{\phi^a_1,\phi^b_1\}=\{\phi^a_1,T^b_2\}=0, \\
\{T^a_2(x_1),T^b_2(x_2)\}=-igf^{abc}T^c_2(x_1)\de(x_1-x_2), \\
\{\phi^a_1,H_0\}=T^a_2, \\
\{T^a_2,H_0\}=igA^b_0f^{bac}T^c_2.
\end{eqnarray}
The extended Hamiltonian takes the form,
\begin{eqnarray}\label{50}
\ti H=\int d^3x \Big[ \frac{1}{2}(\ti p^a_i)^2 + \ti p^a_i
\p_iA^a_0- igf^{abc}A^b_0 A^c_i \ti p^a_i
+\frac{1}{4}(F^a_{ij})^2+ \cr (s^2)^a(\p_i p^a_i- ig f^{abc}p^b_i
A^c_i)+ (v^2)^a \pi^a_2 + v^a \ti p^a_0 \Big].
\end{eqnarray}
Starting from,
\begin{eqnarray}\label{51}
\dot A^a_i=\{A^a_i,H\}=\ti p^a_i + \p_i A^a_0 -
igf^{abc}A^b_0A^c_i + \cr- \p_i(s^2)^a- ig f^{bac}A^c_i (s^2)^b
\end{eqnarray}
we find,
\begin{equation}\label{52}
\ti p^a_i= \dot
A^a_i-\p_i(A^a_0-(s^2)^a)+igf^{abc}(A^b_0-(s^2)^b)A^c_i.
\end{equation}
Thus, $\ti L$ reads,
\begin{eqnarray}\label{53}
\ti L = \int d^3x\Big\{\frac{1}{2}(\dot
A^a_i-\p_i(A^a_0-(s^2)^a)+i g f^{abc} (A^b_0- (s^2)^b)A^c_i)^2+
\cr -\frac{1}{4}(F^a_{ij})^2 \Big\}.
\end{eqnarray}
The symmetries of $\ti L$ are given by,
\begin{eqnarray}\label{54}
\de_1: \de_1 A^a_i=0; \,\,  \de_1 A^a_0=(\ep^1)^a; \,\,
\de_1(s^2)^a=
(\ep^1)^a; \\
\de_2: \de_2A^a_i=-\p_i(\ep^2)^a-igf^{abc}(\ep^2)^b A^c_i;   \de_2
A^a_0=0; \de_2 (s^2)^a=(\dot \ep^2)^a.
\end{eqnarray}
The symmetries of $L$ are easily restored,
\begin{eqnarray}\label{55}
\de_1 A^a_i + \de_2A^a_i= -\p_i(\ep^2)^a-igf^{abc}(\ep^2)^b A^c_i,
\\
\de_1 A^a_0 + \de_2 A^a_0 =(\ep^1)^a,
\end{eqnarray}
where the $\ep$'s obey,
\begin{eqnarray}\label{56}
(\dot \ep^2)^b+ (\ep^1)^b + igA^a_0 f^{abc}(\ep^2)^c=0 \Rightarrow
\cr (\ep^1)^b= -\p_0(\ep^2)^b- i g f^{bca}(\ep^2)^c A^a_0.
\end{eqnarray}
Defining $(\ep^2)^a \equiv - \xi^a$, we obtain the expected
result,
\begin{equation}\label{57}
A^a_{\mu} \rightarrow A'^{a}_{\mu}= A^a_{\mu}+ D^{ac}_{\mu} \xi^c,
\end{equation}
where $D^{ac}_{\mu}=\de^{ac} \p_{\mu}-ig f^{acb}A^b_{\mu}$ is the
covariant derivative.

\subsection{Local symmetry of converted nonlinear sigma model}

In the work [15], it is discussed a method of conversion of
second-class constraints into the first class ones based on
transformations that involve derivatives of the
configuration-space variables. It is useful for covariant
quantization of a theory and in the context of doubly special
relativity [25], for example. Here we consider the converted
version of the nonlinear sigma-model presented in [15]. The model
is useful for the purposes of this work since, after the
conversion, there are only first-class constraints. So, we look
for local symmetries of the action
\begin{equation}\label{57.1}
S= \int d^4 x [\frac{1}{2}(\p_{\mu}\phi^a)^2-2\p_{\mu}e
\p^{\mu}\phi^a \phi^a+ \lambda ((\phi^a)^2-1)].
\end{equation}
The primary constraints and conjugate momenta are,
\begin{eqnarray}\label{57.2}
\frac{\p \mathcal L}{\p \dot \phi^a}=p_a= \dot \phi^a - 2\dot e
\phi^a; \, \frac{\p \mathcal L}{\p \dot e}=p_e= -2 \dot \phi^a
\phi^a; \, \frac{\p \mathcal L}{\p \dot \lambda} = p_{\lambda}=0.
\end{eqnarray}
The expressible velocities are given by,
\begin{eqnarray}\label{57.3}
\dot e = -\frac{1}{4\phi^2}(2 \phi p + p_e); \,\, \dot
\phi^a=p_a-\frac{\phi^a}{2 \phi^2}(2\phi p + p_e).
\end{eqnarray}

We are using the notation $\phi^a \phi^a = \phi^2$. The
Hamiltonian $H_0$ and complete Hamiltonian $H$ are given by,
\begin{eqnarray}\label{57.4}
H_0=\int d^3x \Big[\frac{1}{2}p^2-\frac{(2\phi p + p_e)^2}{8
\phi^2} +\frac{1}{2}(\p_i \phi^a)^2+ \cr -2\p_ie \p^i\phi^a \phi^a
- \lambda(\phi^2-1) \Big]; \\ H=H_0+ \int d^3x v p_{\lambda},
\end{eqnarray}
where $v$ is the corresponding Lagrange multiplier. The secondary
constraint follows from the consistency condition
$0=\{p_{\lambda}(x_1),H \}$. One finds, $G_2=\phi^2-1=0$. We still
find a tertiary constraint: $0=\{G_2(x_1),H \}=-p_e$. $G_3=p_e=0$.

If we define $\{G_I\}=\{G_1=p_{\lambda},G_2= \phi^2-1, G_3 =
p_e\}$, then the gauge algebra is,
\begin{eqnarray}\label{57.5}
\{G_I,G_J\}=0 \Rightarrow c_{IJ}{}^{K} = 0 \forall I, J , K; \\
\{G_1, H_0\}=G_2  \Rightarrow b_{1}{}^2=1,
b_{1}{}^1=b_{1}{}^3=0;  \\
\{G_2,H_0\}=-G_3 \Rightarrow b_{2}{}^3=-1, b_{2}{}^1 = b_{2}{}^2=0; \\
\{G_3,H_0\}=-\p^i\p_iG_2 \Rightarrow b_{3}{}^2=-\p^i\p_i,
b_{3}{}^1 = b_{3}{}^3=0.
\end{eqnarray}
Note that expressions of the form $\p_iG_2$, $\p^i \p_i G_2$, etc
are consequences of the already obtained constraints. They do not
imply simplification of the dynamical equations. So we adopt the
following point: spatial derivatives of constraints does not give
raise to new constraints. So, the procedure stops at the third
stage.

The extended Hamiltonian takes the form,
\begin{eqnarray}\label{57.6}
\ti H=\int d^3 x \Big[\frac{1}{2}\ti p^2-\frac{(2\phi \ti p + \ti
p_e)^2}{8 \phi^2} +\frac{1}{2}(\p_i \phi^a)^2-2\p_ie \p^i\phi^a
\phi^a+ \cr - \lambda(\phi^2-1) + s^2(\phi^2-1) + s^3\ti p_e + v
\ti p_{\lambda} + v^2\pi_2 + v^3 \pi_3  \Big].
\end{eqnarray}
Starting from,
\begin{eqnarray}\label{57.7}
\dot \phi^a=\{\phi^a,\ti H\}=\ti p^a -\frac{2 \phi \ti p + \ti
p_e}{2 \phi^2}\phi^a \\
\dot e= \{e, \ti H\}=-\frac{2 \phi \ti p + \ti p_e}{4 \phi^2} +
s^3,
\end{eqnarray}
we find,
\begin{equation}\label{57.8}
\ti p_a=\dot \phi^a - 2 \phi^a(\dot e -s^3), \\
\ti p_e = -2\phi \dot \phi.
\end{equation}
Thus, $\ti L$ reads,
\begin{eqnarray}\label{57.9}
\ti L = \int d^3x\Big\{\frac{1}{2}(\p_{\mu}\phi^a)^2 -2\phi \dot
\phi(\dot e -s^3)+ 2\p_ie \p^i \phi^a \phi^a+ \cr + (\lambda
-s^2)(\phi^2-1) \Big\}.
\end{eqnarray}
The symmetries of $\ti L$ are given by,
\begin{eqnarray}\label{57.10}
\de_1: \de_1 \phi^a=0;   \de_1 \lambda=\ep^1;
\de_1 e=0;  \de_1s^2= \ep^1;  \de_1 s^3=0.   \\
\de_2: \de_2 \phi^a=0;   \de_2 \lambda=0;  \de_2 e=0;
\de_2s^2=\dot \ep^2;  \de_2 s^3=-\ep^2. \\
\de_3: \de_3 \phi^a=0;  \de_3 \lambda=0; \de_1 e=\ep^3;
\de_3s^2=b_{3}{}^2 \ep^3 = -\p_i\p_i \ep^3; \de_3 s^3=\dot \ep^3.
\end{eqnarray}
The symmetries of $L$ are restored,
\begin{eqnarray}\label{57.11}
\de_1 \phi^a + \de_2 \phi^a + \de_3 \phi^a=0,
\\
\de_1 \lambda + \de_2 \lambda + \de_3 \lambda =\ep^1, \\
\de_1 e + \de_2 e + \de_3 e =\ep^3,
\end{eqnarray}
where the $\ep$'s obey,
\begin{eqnarray}\label{57.12}
\dot \ep^2 -\p_i \p_i \ep^3 + \ep^1 = 0, \\
\dot \ep^3- \ep^2=0.
\end{eqnarray}
Defining $\ep^3 \equiv - \ep$, we obtain the following local
symmetry,
\begin{eqnarray}\label{57.13}
\de \phi^a = 0; \,\, \de \lambda = \p_{\mu} \p^{\mu}\ep; \,\, \de
e= - \ep.
\end{eqnarray}
where $\ep = \ep (x)$ is an arbitrary function of space-time
coordinates.

\section{Conclusion}
In this work we have presented a generalization of the extended
Lagrangian method of finding local symmetries to the field
systems. As we have illustrated on various examples, it provides a
systematic method of finding gauge symmetries of a singular
Lagrangian $L$ with first-class constraints. The initial theory is
deformed in a special way such that all the symmetries of the
deformed Lagrangian $\tilde L$ can easily be found. The symmetries
of $L$ are also obtained. According to the scheme, all the
first-class constraints of the initial theory are the gauge
generators of the deformed theory. We also pointed out the
subtleties that must be taken into account when moving from
classical systems to the continuous case. In this context we
briefly discussed some fundamental definitions that are slightly
complicated and do not follow directly from point mechanics to
field theories, including degrees of freedom and constraints.

\section*{Acknowledgments}
B. F. R. would like to thank the Brazilian foundation FAPEAM -
Programa RH Interiorização - for financial support.

\end{document}